\documentclass[conference,peer-reviewed]{aesconf_new}
\graphicspath{{./}{figures/}}

\usepackage[utf8]{inputenc}

\usepackage{microtype}

\usepackage[numbers,square]{natbib}

\usepackage{booktabs}
\usepackage{color}
\usepackage{url}
\usepackage{booktabs} 
\usepackage{multirow} 
\usepackage{adjustbox}

\title{Integrating IP broadcasting with audio tags: Workflow and challenges}

\author[2]{Rhys Burchett-Vass}
\author[1,2]{Arshdeep Singh}
\author[1,2]{Gabriel Bibbó}
\author[1,2]{Mark D. Plumbley}
\affil[1]{Centre for Vision Speech and Signal Processing (CVSSP)}
\affil[2]{University of Surrey, UK}

\correspondence{Arshdeep Singh}{arshdeep.singh@surrey.ac.uk}

\lastnames{Burchett-Vass, Singh, Bibbó, and Plumbley}

\shorttitle{Integrating IP broadcasting with audio tags}

\begin{document}

\twocolumn[
\maketitle 

\begin{onecolabstract}
The broadcasting industry 
has adopted IP technologies, revolutionising both live and pre-recorded content production, from news gathering to live music events. IP broadcasting allows for the transport of audio and video signals in an easily configurable way, aligning with modern networking techniques. This shift towards an IP workflow allows for much greater flexibility, not only in routing signals but with the integration of tools using standard web development techniques. One possible tool could include the use of live audio tagging, which has a number of uses in the production of content. These could include adding sound effects to automated closed captioning or identifying unwanted sound events within a scene. In this paper, we describe the process of containerising an audio tagging model into a microservice, a small segregated code module that can be integrated into a multitude of different network setups. The goal is to develop a modular, accessible, and flexible tool capable of seamless deployment into broadcasting workflows of all sizes, from small productions to large corporations. Challenges surrounding latency of the selected audio tagging model and its effect on the usefulness of the end product are discussed.
\end{onecolabstract}
]

\section{Introduction}
The audio track contains a wide array of descriptive information about the sound events in a scene. Detecting sound events in real-time could have a number of uses in the broadcasting industry, from aiding operators in programme creation to enhancing end user accessibility. 
For example, BBC Research and Development \cite{BBC_audiowatch} employed a sound event detection framework to identify sounds that may disrupt the ambience of a program. This work was targeted at the BBC's \textit{Autumnwatch} programme which uses wildlife cameras capturing the movement of animals. To avoid undesirable noises interrupting the live stream, such as cars passing by or people talking, an icon is overlayed onto the operator's interface, indicating the undesirable sound event so the operator does not to switch to that source. Another use of sound event detection is closed captioning. While the majority of the existing work in captioning broadcast audio has been focused on analysing the speech events \cite{raimond2012automated, levin2014automated}, only a few attempts of identifying sound events in a real time (live) transmission has been conducted \cite{BBC_audiowatch}. To achieve full captioning (closed captions of both sound events and speech) within IP broadcasting, general sound event detection models that detect sound events including speech are required.

Internet Protocol (IP) broadcasting describes the process of transmitting audio and video signals from one location to another using IP networking. One approach traditionally used for transmitting audio/video is Serial Digital Interface (SDI), with fixed connections between dedicated hardware devices.
IP broadcasting allows software to replace some of these hardware devices, enabling greater scalability and easy re-configuration. Cloud technology and containerisation methods such as Docker \cite{docker_2022} can be utilised to take advantage of such scalability. 

There are a few challenges in creating software for use in an IP broadcasting environment.
First, as with most modern web applications, is scalability and containerisation of software, which allows the infrastructure to adapt depending on the demand on the system by starting up new containers when required. Containerisation also allows for the same task to be conducted independently on different streams or sources by having a container per stream. Another advantage of containerisation is that, if a fault occurs on one container, it does not damage the entire system as a whole and can be fixed independently. The second challenge is handling of the inelastic audio and video traffic, which does not adapt well to changes in network conditions, without introducing delay and jitter to the transmission.

\begin{figure}[t]
    \centering
    \includegraphics[ width=0.5\textwidth]{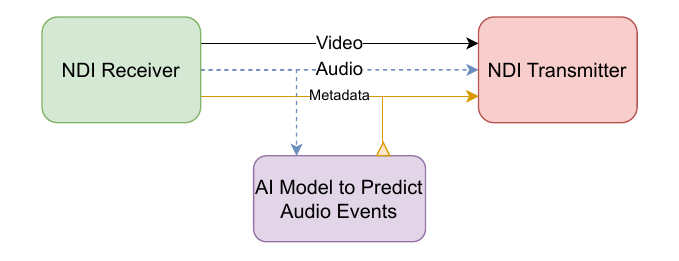}
    \caption{Basic flow of audio, video and metadata frames travelling through their respective data streams using Network Device Interface (NDI) system.}
    \label{fig:panns_module_block}
\end{figure}

To overcome above challenges and integrate sound event data into IP broadcasting, this paper contributes:
\begin{itemize}
    \item Containerising audio tagging applications to isolate each component from the other elements of the system, i.e. other processing units, transmission and reception code. Code isolation allows the component to run on any machine. Containerisation makes it easy to create multiple instances of a component if needed, allowing for scalability and the use of cloud platforms.
    \item Leveraging an Artificial Intelligence (AI) model to generate audio tags to transmit meta-information alongside audio. This added information about the contents of the audio track has a number of uses in the area of automation - from better production tools to improved accessibility via more descriptive automated captioning.  An overall proposed framework is shown in Figure \ref{fig:panns_module_block} and  more details  including challenges around our approach can be found  in Section \ref{sec:challenges}.  Our codes are available at Github\footnote{\url{https://github.com/Rhysbv/panns_ndi}}.
\end{itemize}


\section{Related work}
\label{sec: related work}
\subsection{ A Brief Overview on IP Broadcasting Technology}
\label{sec:related_work:broadcast}
There are a few technologies currently used for IP broadcasting. The first of these are described in standards from the Society of Motion Picture and Television Engineers (SMPTE) as the ST 2110 suite of standards \cite{smpte2019OV21100}. SMPTE standards are used by the industry with examples including the Serial Digital Interface (SDI) standards for transmission between equipment over a direct connection, i.e. coaxial or fibre optic cable. Secondly, the Networked Media Open Specifications (NMOS) from the Advanced Media Workflow Association (AMWA) uses ST 2110 along with other standards to define APIs allowing for the connection of multiple receivers and senders on a network in a vendor agnostic way. NMOS is not software, but a set of specifications aiding development in the software. On the other hand, Network Device Interface (NDI) by NewTek \cite{newtek_ndi} 
is an open standard with fully developed software and Software Development Kit (SDK), designed to allow for easy integration of IP broadcasting into existing software by utilising the NDI SDK.  

\subsection{Audio Recognition in Broadcasting}
\label{sec:related_work:audio_rec}
There has been some work conducted in broadcasting related to recognition of audio events \cite{BBC_audiowatch}. However most examples relate to speech recognition and transcription, commonly used for tagging content for archiving purposes.
Recognising speech allows for easy searching without having to manually tag content. For example, Raimond et al. \cite{raimond2012automated} describe a system to automate the tagging of content within the BBC's radio archive based on speech audio. Levin et al. \cite{levin2014automated} describe a system using automatic speech recognition for captioning of news programming. This system  runs against a re-speaker, a person repeating speech in a more readable and understandable way for the system, avoiding the issues surrounding the acoustic environment and overlapping speakers. However, this system  only supports the processing of speech and does not consider sound events in general. More modern proprietary solutions do exists \cite{idm_closed, encaption} which remove the requirement for a re-speaker but are still incapable of including sound events. Additionally, BBC Research and Development \cite{BBC_audiowatch} have designed an application program (a software tool) to identify sound events for the purpose of audio monitoring. 

\section{System Design}
\label{sec:design}

In contrast to previous work, we separate the audio tagging software from any other application programs. In our work, a modular approach is opted for and a container specifically for general audio tagging is built to allow multiple applications on the network to take advantage of the technology without repeating work. This is helpful considering the computational overhead associated with AI models. Our system can include the environment audio event monitoring system 
as described by BBC Research and Development \cite{BBC_audiowatch}  in addition to other systems e.g. for captioning. 

\label{sec:chosen_tech}
\subsection{Selecting IP Broadcasting Approach}
For our work, we need an IP based technology that is both well used by the industry and is simple to implement, allowing for wide adoption of the audio tagging technology.
Standards have been created to support new IP based workflows. One example discussed in Section \ref{sec:related_work:broadcast} is the ST 2110 suite of standards authored by SMPTE. These describes the transport of compressed and uncompressed audio, video and metadata via individual Real-time Transport Protocol (RTP) streams. However, the complexity of understanding the suite of standards means that it is only practical for large corporations to implement. An alternative standard mentioned in Section \ref{sec:related_work:broadcast} is the Network Device Interface (NDI), an open standard created by NewTek \cite{newtek_ndi} which has an easy to use Software Development Kit (SDK). 
NDI is available in a wide variety of software and hardware applications, enabling its wide spread adoption in both large and small operations. This has lead to NDI being the selected approach for our work. NDI transports data in the form of frames that contain the relevant data as well as supporting information to aid in its use. There are three types of frames used by NDI: audio, video and metadata (Figure \ref{fig:panns_module_block}). NDI also handles the detection of sources allowing for routing of NDI frames.

\subsection{AI Model used for Audio Tagging}
To identify audio tags, we use convolutional neural networks (CNNs), that have shown good performance in many audio classification tasks \cite{purwins2019deep, kong2020panns}. For example, pre-trained audio neural networks (PANNs) \cite{kong2020panns} have been widely used to recognize a variety of audio events. A description of the CNN models used in this paper for predicting the audio tags is given below.

\textbf{Pre-trained Audio Neural Networks (PANNs): }
CNN14 \cite{kong2020panns} is a pre-trained audio neural network  that is trained on Google AudioSet dataset \cite{gemmeke2017audio}. CNN14 is trained  by extracting the log-mel spectrograms from the audio clips. CNN14 has 81M parameters and it takes 21G multiply-accumulate operations (MACs) to predict tags of the audio of length 10 seconds. The trained CNN14  can predict wide range of sound events such as car passing by, speech, siren, animal etc. This helps identifying sounds in the wide array of possible scenarios the system could be exposed to, i.e. different types of broadcast programming such as news gathering in various locations or a panel show within a studio.

\textbf{Efficient PANNs: } E-PANNs \cite{singh2023panns} is an efficient version of original PANNs with reduced memory requirement  (24M parameters) and a reduced computational complexity (13G MACs per 10 seconds audio). 
The efficient AI models are beneficial in an IP networking environment, especially one involving inelastic traffic (network traffic that is sensitive to variations in delay, e.g. audio and video streams). This will be explored in Section \ref{sec:challenges}.

\subsection{NDI Integration}
We use the NDI SDK \cite{NDISDK6} to create a software module including the PANNs algorithm. Due to the reliance on Python based packages within the PANNs module such as \textbf{PANNs inference} \cite{qiuqiangkong_qiuqiangkongpanns_inference_2024}, we use Python for implementation.  Specifically, community Python bindings  \cite{kondo_buresundi-python_2023} to interface between Python and the C++ SDK are used to enable NDI support. An additional Python package is created to simplify the process of integrating NDI into both the PANNs module and potential proof of concept applications described in Section \ref{sec:workflow}. 
Our Python package contains three classes: A receiver, transmitter and finder as seen in Figure \ref{fig:panns_module_block}. An application can find NDI network sources using the finder class. Frames can be received from an NDI source using the receiver class. Received Frames can then be processed and transmitted by creating its own NDI source using the transmitter class. An example of how this is used here can be seen in Figure \ref{fig:panns_module_block}.
The flow of audio, video and metadata frames is uninterrupted between the receiver and transmitter. Each audio frame is intercepted and a copy is taken for analysis while the original copy is sent straight to the transmitter, minimising delay and jitter. 
One issue surrounding the community supplied Python bindings were the associated bugs, especially surrounding memory management. This led to having to convert each frame to a Python dataclass so that it could be effectively freed and dealt with by the Python garbage collector, an issue that would not have been encountered using the original C++ SDK.

\subsection{Integrating Sound Events Metadata}
\begin{figure*}[t]
    \centering
    \includegraphics[ width=0.9\textwidth]{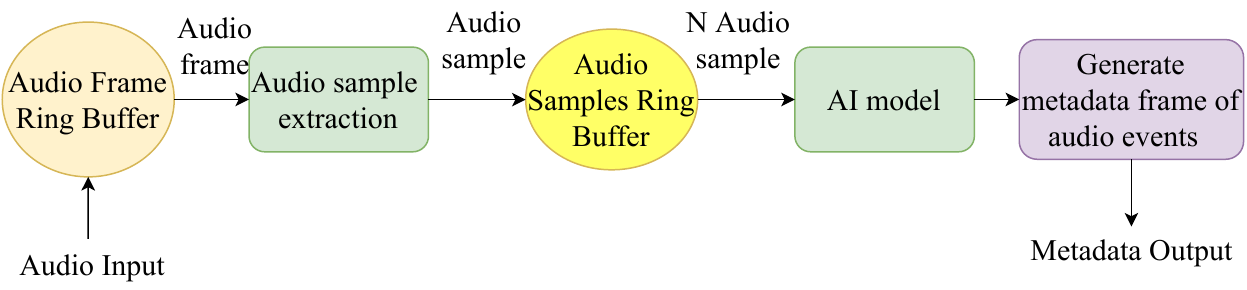}
    \caption{Metadata generation pipeline.}
    \label{fig:pannsflow}
\end{figure*}

In order for E-PANNs to produce sound event predictions from the PANNs model and make it compatible with other NDI applications, we follow the pipeline as shown in Figure  \ref{fig:pannsflow} that takes the incoming audio frames from NDI and creates metadata frames containing the audio tag to be sent across the network.  

We use two ring buffers. The first ring buffer stores incoming audio frames. From each audio frame, we extract the individual audio samples and store them in a second ring buffer. Once a sufficient number of samples has been collected in the second ring buffer the entire contents of the ring buffer is fed into the PANNs model. The size of the second ring buffer determines the duration of the audio window that PANNs analyses. The impact of the size of the window on the models latency is discussed in Section \ref{sec:challenges}. 
To distribute the predicted sound event across the NDI network, we use metadata frames. These frames transport XML data, which can include third-party metadata as used here. The output text from PANNs is inserted into an XML template for transmission. Other NDI applications can then receive this XML via the metadata frames to access the sound event prediction.

A summary of various steps is explained below:

\begin{enumerate}
    \item Store received audio frames in the first ring buffer.
    \item Extract the floating point Pulse Code Modulated (PCM) audio samples from each frame and store these in the second ring buffer.
    \item Wait until a given number of samples have been collected.
    \item Feed the entire contents of the second ring buffer into the CNN model.
    \item Generate a metadata frame containing the prediction from the CNN model.
\end{enumerate}

\begin{figure*}[ht]
    \centering
    \includegraphics[ width=0.9\textwidth]{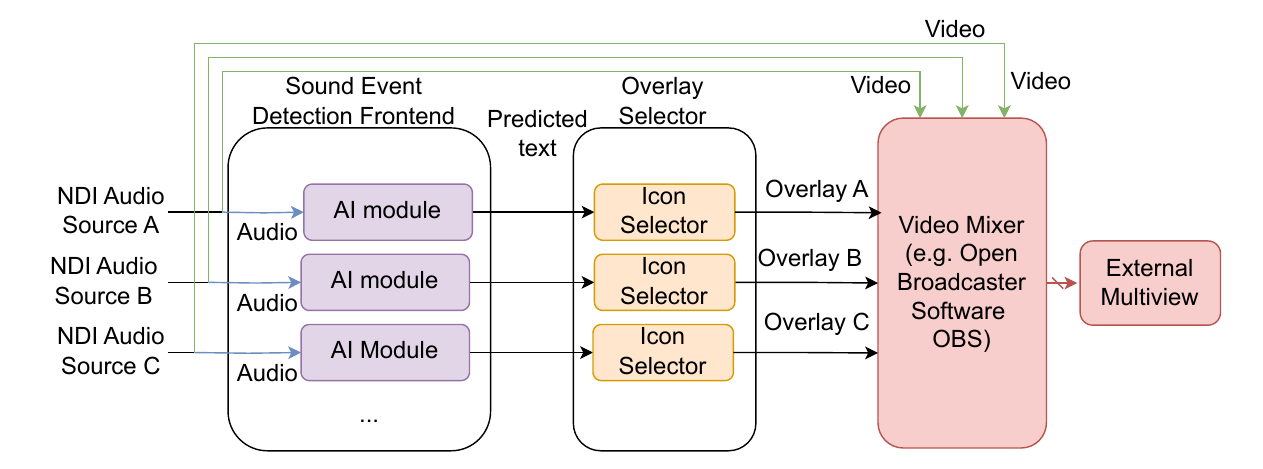}
    \caption{Proposed integrated pipeline: Audiowatch \cite{BBC_audiowatch} framework with a separated audio tagging unit.}
    \label{fig:audiowatch}
\end{figure*}

\section{Example Workflow}
\label{sec:workflow}

The proposed containerised component allows for the integration of audio tagging capabilities into a multitude of different systems and use cases.  Below, we provide two examples of integration of audio tagging system into existing IP broadcasting framework,

\subsection{Audiowatch Example}
\label{sec:workflow:audiowatch}

Figure \ref{fig:audiowatch} demonstrates 
a pipeline inspired by the BBC Audiowatch project, showing an audio tagging module separated from other application programs
We use Docker \cite{docker_2022} for containerisation, creating multiple instances of the audio tagging software to analyse several NDI sources simultaneously.
A sound event detection front end is a dashboard user interface as shown in Figure \ref{fig:pannsflow} and it  generates metadata corresponding to input audio. Metadata containing sound event information is then sent to the icon selector module for processing.   
Next, various icon selector containers extract the sound events from the audio track supplied within the metadata frames. After identifying the unwanted sound events, an appropriate icon overlay is transmitted as an NDI video frame. Next, a video mixing software tool such as Open Broadcaster Software (OBS) \cite{obsprojectobs-studio_2023} is used to superimpose the icon onto the original video source for displaying on the operators multiview, which is used to monitor all video sources.

\subsection{Online Closed Captioning}
Another example integration could be the use of audio tagging to enhance closed captioning. As discussed in Section \ref{sec:related_work:audio_rec}, while work has been conducted to automated closed captioning in real time using automatic speech recognition, these do not include descriptions of sound events. By combining the two technologies, full closed captioning could be achieved. This would involve first parsing the audio through the audio tagging model using our container. When the result is returned as human speech, the audio would then be passed through a second speech recognition model to generate subtitles. One major concern would be the accuracy of the audio tagging model. If the speech was not always detected, we would miss large portions of speech text. Additionally the difference in latency between a sound event being inserted and speech going through two models would have to be accounted for.

\begin{figure*}[t]
    \centering
    \includegraphics[scale=0.38]{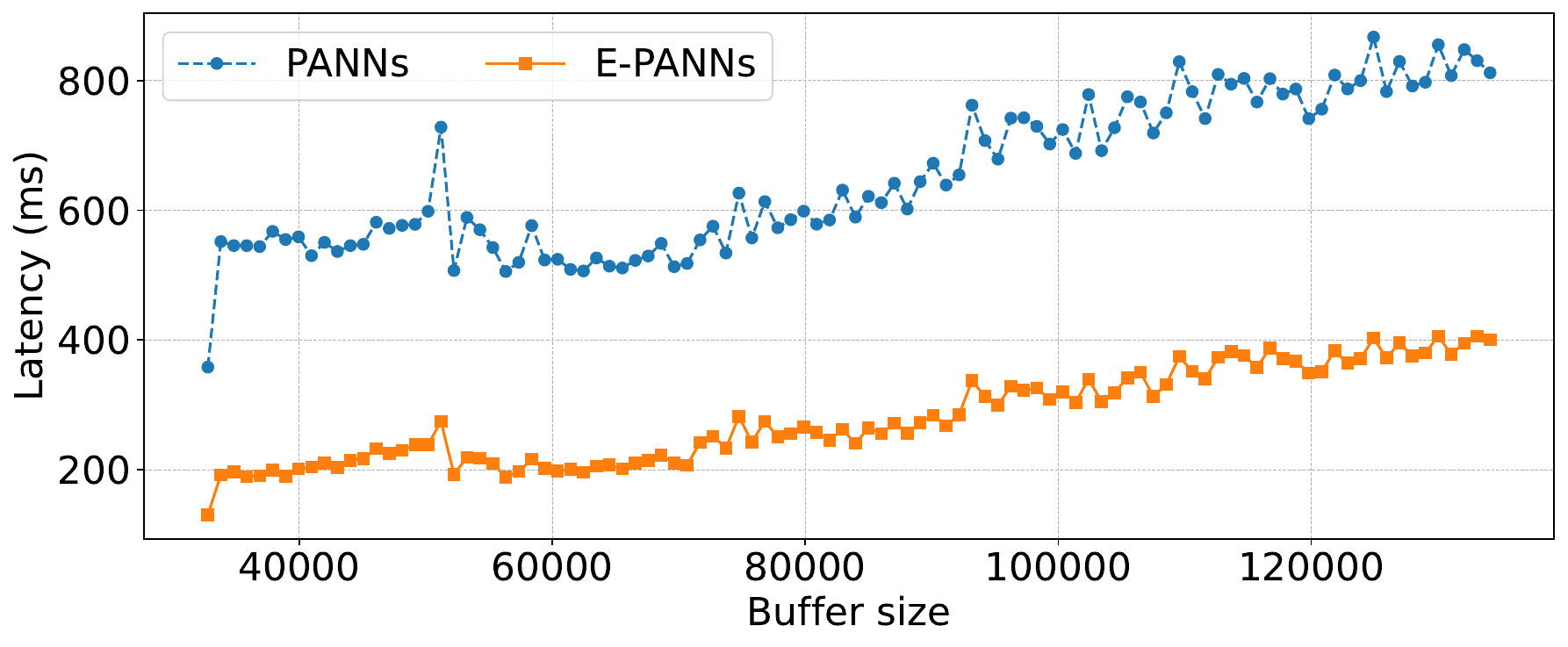}
    \caption{PANNs/E-PANNs model latency vs buffer size when inputting audio sampled at 48KHz. Experiments are performed a on AMD Ryzen 5 2500U system at 2GHz.}
    \label{fig:Latency_graph}
\end{figure*}

\section{AI model Integration Challenges}
\label{sec:challenges}
There are a number of integration challenges to consider while designing AI based software fit for broadcasting. These challenges include the accuracy of the prediction and the latency of the model delaying the signal. Generally, PANNs and E-PANNs give similar prediction results.

\textbf{Model latency:}
The latency of the model here describes the amount of time it takes given a number of audio samples to produce an accurate sound event prediction. Consideration of the model latency is significant given that we are dealing with inelastic audio and video network traffic. This means that any delay in processing contributes to a delay in the resulting transmission depending on the infrastructure. The delay can be mitigated using a design similar to that shown in Figure \ref{fig:panns_module_block}, however there is the issue of predictions being 
desynchronised from events heard in the audio track 
Although we have minimal control over the IP network using the audio tagging module, and thus cannot manage the network's latency, we can still select an optimal model that minimises latency while maintaining accuracy.

\textbf{Buffer size versus model latency:}
To analyse buffer size and latency of model, we performed experiments using a set of audio recordings with known sound events. 
The first audio recording is taken directly from the PANNs repository. It 
involves a telephone ringing followed by human speech and is of seven seconds. 
The second audio recording of a car driving into the distance. The third audio recording is created by mixing a car driving and a running river sound events. 

Given the audio recordings, we analyse the latency of the CNN model at different number of audio samples. We generate different length audio segments. The audio samples taken are of multiples of 1024 (assuming frames containing 1024 samples are used) and represents the size of the buffer.  Given the audio samples of different length, we use the PANNs or E-PANNs model to produce predictions while measuring the time taken for the model to produce a prediction. Figure \ref{fig:Latency_graph} shows the latency of the PANNs and E-PANNs models at different buffer size. Both PANNs and E-PANNs follow a similar trajectory, with E-PANNs showing a considerable improvement in latency. This suggests that choosing an appropriate model contributes to try to improve latency and hence making integration of audio events more real-time while using less resources.

Experiments found that in our example a buffer size of 47 frames with 1024 samples each
This equates to an audio window with a duration of 1.002s sampled at 48KHz, that gives correct results with the minimal latency. Model latency computed on an AMD Ryzen 5 2500U and Intel Core i9-13900HX hardware can be found in Figure \ref{fig:Latency_graph} and prediction results can be found in Table \ref{tab:comparison}.

\begin{table}[h]
    \centering
    \caption{Comparison of PANNs and E-PANNs predictions across different buffer sizes}
    \label{tab:comparison}
    \small 
    \setlength{\tabcolsep}{5pt} 
    \renewcommand{\arraystretch}{0.7} 
    \begin{adjustbox}{max width=0.5\textwidth} 
    \begin{tabular}{cccc}
        \toprule
        \textbf{Buffer Size} & \textbf{Groundtruth} & \textbf{PANNs} & \textbf{E-PANNs} \\
        \midrule
        32768  & Tone           & Sine wave              & Sine wave \\
               & Car            & Music                  & Music \\
               & River \& Car   & Waves, surf            & Boat, water vehicle \\
               & Phone \& Voice & Telephone bell ringing & Telephone bell ringing \\
        \midrule
        38912  & Tone           & Sine wave              & Sine wave \\
               & Car            & Music                  & Music \\
               & River \& Car   & Vehicle                & Boat, water vehicle \\
               & Phone \& Voice & Music                  & Telephone bell ringing \\
        \midrule
        48128  & Tone           & Sine wave              & Sine wave \\
               & Car            & Silence                & Vehicle \\
               & River \& Car   & Vehicle                & Vehicle \\
               & Phone \& Voice & Music                  & Telephone bell ringing \\
        \midrule
        56320  & Tone           & Sine wave              & Sine wave \\
               & Car            & Vehicle                & Vehicle \\
               & River \& Car   & Vehicle                & Vehicle \\
               & Phone \& Voice & Music                  & Telephone bell ringing \\
        \midrule
        61440  & Tone           & Sine wave              & Sine wave \\
               & Car            & Vehicle                & Vehicle \\
               & River \& Car   & Vehicle                & Vehicle \\
               & Phone \& Voice & Telephone bell ringing & Telephone bell ringing \\
        \midrule
        65536  & Tone           & Sine wave              & Sine wave \\
               & Car            & Vehicle                & Vehicle \\
               & River \& Car   & Vehicle                & Vehicle \\
               & Phone \& Voice & Telephone bell ringing & Telephone bell ringing \\
        \midrule
        102400 & Tone           & Sine wave              & Sine wave \\
               & Car            & Vehicle                & Vehicle \\
               & River \& Car   & Waves, surf            & Vehicle \\
               & Phone \& Voice & Telephone bell ringing & Telephone bell ringing \\
        \midrule
        134144 & Tone           & Sine wave              & Sine wave \\
               & Car            & Vehicle                & Vehicle \\
               & River \& Car   & Vehicle                & Vehicle \\
               & Phone \& Voice & Speech                 & Speech \\
        \bottomrule
    \end{tabular}
    \end{adjustbox}
\end{table}

\section{Discussion and Conclusion}
\label{sec:conclusion}

The integration of IP broadcasting with audio tagging offers significant potential for enhancing broadcast workflows, but it also presents several challenges. The transition to IP broadcasting enables a more flexible, scalable, and reconfigurable infrastructure compared to traditional methods based on Serial Digital Interface (SDI). This flexibility is further enhanced by containerisation technologies making the system more resilient and adaptable. However, implementing an audio tagging system introduces challenges primarily related to latency and the accuracy of audio tagging models.

One of the primary challenges discussed is the latency associated with the audio tagging model. Given the real-time nature of broadcasting, any delays introduced by processing can impact the overall operation. This makes the choice of buffer size crucial. A smaller buffer reduces latency but might compromise the accuracy of sound event detection. Conversely, a larger buffer improves accuracy but increases latency. Experiments conducted (Table \ref{tab:comparison}) show that an acceptable balance can be achieved with a buffer size of 48128 samples, which provides an acceptable latency while maintaining accuracy. The use of Efficient PANNs (E-PANNs) further helps in reducing the computational complexity and memory requirements, making it a suitable choice for real-time applications.

Containerisation offers a robust solution to scalability issues. By isolating the audio tagging functionality into a microservice, it becomes possible to scale the system by simply adding more containers as needed. This isolation also ensures that a fault in one container does not affect the entire system, enhancing overall reliability. The use of Docker to containerise these services allows for easy deployment and management across different network setups. Additionally, the integration with NDI
, which is widely adopted in the industry, ensures broad applicability.

Despite these advantages, real-world deployment of such a system is not without hurdles. The reliance on Python bindings to interface with the NDI SDK, while practical, introduces potential issues with memory management that need careful handling.

\subsection{Conclusion}
Integrating IP broadcasting with audio tagging presents a promising advancement for the broadcasting industry. The use of containerisation and audio tagging for real-time sound event detection can significantly enhance content production and accessibility. However, addressing the challenges of latency, accuracy and real-world deployment is crucial for the successful implementation of this approach. 
Future work includes re-writing the codebase to use the NDI C++ SDK directly, avoiding the issues surrounding the community Python bindings. Additionally, we would like to analyse more complex models such as transformers \cite{gong2021ast, schmid2023efficient}  within our broadcasting framework. Finally, the creation of the discussed proof of concept applications would allow for full demonstration of the usefulness of this technology.

\section{Acknowledgements}

This work was supported by the Engineering and Physical Sciences Research Council [grant number EP/T019751/1] and an Adobe Research Gift. For the purpose of open access, the authors have applied a Creative Commons Attribution (CC BY) licence to any Author Accepted Manuscript version arising.

\bibliographystyle{jaes}

\bibliography{refs}

\begin{thebibliography}{18}
\newcommand{\enquote}[1]{``#1''}
\providecommand{\natexlab}[1]{#1}
\expandafter\ifx\csname urlstyle\endcsname\relax
  \providecommand{\doi}[1]{doi:\discretionary{}{}{}#1}\else
  \providecommand{\doi}{doi:\discretionary{}{}{}\begingroup \urlstyle{rm}\Url}\fi

\bibitem[{Ward and Dawes(2021)}]{BBC_audiowatch}
Ward, S. and Dawes, R., \enquote{{AudioWatch} - Live audio monitoring for Autumnwatch 2021 - {BBC} R\&D,} 2021.

\bibitem[{Raimond et~al.(2012)Raimond, Lowis, Hodgson, and Tweed}]{raimond2012automated}
Raimond, Y., Lowis, C., Hodgson, R., and Tweed, J., \enquote{Automated semantic tagging of speech audio,} in \emph{Proceedings of the 21st International Conference on World Wide Web}, pp. 405--408, 2012.

\bibitem[{Levin et~al.(2014)Levin, Ponomareva, Bulusheva, Chernykh, Medennikov, Merkin, Prudnikov, and Tomashenko}]{levin2014automated}
Levin, K., Ponomareva, I., Bulusheva, A., Chernykh, G., Medennikov, I., Merkin, N., Prudnikov, A., and Tomashenko, N., \enquote{Automated closed captioning for Russian live broadcasting,} in \emph{Fifteenth Annual Conference of the International Speech Communication Association}, 2014.

\bibitem[{Docker(2022)}]{docker_2022}
Docker, \enquote{Docker: Accelerated Container Application Development,} 2022, https://www.docker.com/.

\bibitem[{SMPTE(2018)}]{smpte2019OV21100}
SMPTE, \enquote{{SMPTE} {OV} 2110-0:2018 - {SMPTE} Overview Document - Professional Media over Managed {IP} Networks Roadmap for the 2110 Document Suite,} 2018, https://cta-redirect.hubspot.com/cta/redirect/5253154/c8fbb67b-5cba-421d-ad95-29de41239ef7.

\bibitem[{{NewTek}(2023)}]{newtek_ndi}
{NewTek}, \enquote{{NDI} 5.6 White Paper,} Technical report, NewTek, 2023, https://ndi.video/wp-content/uploads/2023/09/NDI-5.6-White-Paper-2023.pdf.

\bibitem[{IBM(2025)}]{idm_closed}
IBM, \enquote{Closed Captioning Software - {IBM} Watson Media,} 2025, https://www.ibm.com/consulting/closed-captioning.

\bibitem[{enc(2025)}]{encaption}
\enquote{{enCaption}: Automated Closed Captioning System {\textbar} {ENCO} Systems,} 2025.

\bibitem[{Purwins et~al.(2019)Purwins, Li, Virtanen, Schl{\"u}ter, Chang, and Sainath}]{purwins2019deep}
Purwins, H., Li, B., Virtanen, T., Schl{\"u}ter, J., Chang, S.-Y., and Sainath, T., \enquote{Deep learning for audio signal processing,} \emph{IEEE Journal of Selected Topics in Signal Processing}, 13(2), pp. 206--219, 2019.

\bibitem[{Kong et~al.(2020)Kong, Cao, Iqbal, Wang, Wang, and Plumbley}]{kong2020panns}
Kong, Q., Cao, Y., Iqbal, T., Wang, Y., Wang, W., and Plumbley, M.~D., \enquote{PANNs: Large-scale pretrained audio neural networks for audio pattern recognition,} \emph{IEEE/ACM Transactions on Audio, Speech, and Language Processing}, 28, pp. 2880--2894, 2020.

\bibitem[{Gemmeke et~al.(2017)Gemmeke, Ellis, Freedman, Jansen, Lawrence, Moore, Plakal, and Ritter}]{gemmeke2017audio}
Gemmeke, J.~F., Ellis, D.~P., Freedman, D., Jansen, A., Lawrence, W., Moore, R.~C., Plakal, M., and Ritter, M., \enquote{Audio Set: An ontology and human-labeled dataset for audio events,} in \emph{IEEE International Conference on acoustics, Speech and Signal Processing (ICASSP)}, pp. 776--780, 2017.

\bibitem[{Singh et~al.(2023)Singh, Liu, and Plumbley}]{singh2023panns}
Singh, A., Liu, H., and Plumbley, M.~D., \enquote{E-PANNs: Sound Recognition Using Efficient Pre-trained Audio Neural Networks,} in \emph{INTER-NOISE and NOISE-CON Congress and Conference Proceedings}, volume 268, pp. 7220--7228, Institute of Noise Control Engineering, 2023.

\bibitem[{NewTek(2024)}]{NDISDK6}
NewTek, \enquote{NDI SDK V6.0.1,} 2024, https://ndi.video/for-developers/ndi-sdk.

\bibitem[{kong(2020)}]{qiuqiangkong_qiuqiangkongpanns_inference_2024}
kong, Q., \enquote{qiuqiangkong/panns\_inference,} 2020, original-date: 2020-03-08T06:22:30Z.

\bibitem[{Kondo(2022)}]{kondo_buresundi-python_2023}
Kondo, N., \enquote{{NDI-Python} bindings,} 2022, https://github.com/buresu/ndi-python.

\bibitem[{Project(2025)}]{obsprojectobs-studio_2023}
Project, O., \enquote{OBS studio project,} 2025, https://github.com/obsproject/obs-studio.

\bibitem[{Gong et~al.(2021)Gong, Chung, and Glass}]{gong2021ast}
Gong, Y., Chung, Y.-A., and Glass, J., \enquote{AST: Audio spectrogram transformer,} \emph{Interspeech}, 2021.

\bibitem[{Schmid et~al.(2023)Schmid, Koutini, and Widmer}]{schmid2023efficient}
Schmid, F., Koutini, K., and Widmer, G., \enquote{Efficient large-scale audio tagging via transformer-to-{CNN} knowledge distillation,} in \emph{IEEE International Conference on Acoustics, Speech and Signal Processing (ICASSP)}, IEEE, 2023.

\end{thebibliography}

\end{document}